\documentclass[12pt]{iopart}
\usepackage{iopams}  
\usepackage{graphicx}
\usepackage{sidecap}
\usepackage{caption}
\begin{document}

\title [Materials Physics and Spin Glasses ] {Materials Physics and Spin Glasses \footnote{This paper is dedicated to Giorgio Parisi in celebration of his 70th birthday,  ‍in   appreciation of his enormously innovative and influential contributions  to science and  its leadership and in gratitude for his friendship.} }

\author{David Sherrington}

\address{Rudolf Peierls Centre for Theoretical Physics, Clarendon Laboratory, Parks Rd., Oxford  OX1 3PU, UK

Santa Fe Institute, 1304 Hyde Park Rd., Santa Fe, NM  87501, USA}
\ead{david.sherrington@physics.ox.ac.uk}
\vspace{10pt}
\begin{indented}
\item[]{12 February 2019}
\end{indented}

\begin{abstract}

Comparisons and analogies are drawn between materials ferroic glasses and conventional spin glasses,  in terms of both experiment and theoretical modelling, with inter-system conceptual transfers leading to suggestions of further issues to investigate.

\end{abstract}

\section{Introduction}

The original physical systems that led to the designation `spin glass' 
were substitutional alloys of non-magnetic and magnetic metals, such as {\bf{Au}}Fe and {\bf{Cu}}\rm{Mn}, which exhibited unusual features at low temperatures \cite{Mydosh}.

The challenge to understand them led to the recognition of a need for and the formulation of new statistical physics, new concepts and new methodology, particularly driven by the concept of replica symmetry breaking and its sophisticated formulation and physical understanding by Giorgio Parisi \cite{Parisi1, Parisi2}. 
This theoretical study, in turn, has stimulated extensions of conceptualization, mathematical formulation and practical application in many other areas of many body physics, probablity theory, computer science, biology and econophysics
\cite{MPV, Nishimori},
typically characterized by frustrated interactions and quenched (or slower evolving) disorder. It continues to lead to new advances, often in areas which were not anticipated when the original materials were studied and typically in situations where the systems and their interacting microscopic actors are physically different from those in the original metallic alloys. 
These theoretical  extensions and extrapolations have had both fundamental and practical importance. 

However, the original spin glass materials themselves have their onset phase transitions at rather low temperatures and have not  had  practical application. The purpose of this brief paper is to highlight some other material systems which were discovered decades ago to have interesting and applicable characteristics 
yet lack a generally agreed understanding, and to argue that they are essentially soft pseudo-spin glasses that also pose some  conceptual issues  for  statistical physics theory.
 
 \section{Relaxors}
 
\begin{figure*}
\includegraphics*[width=.3150\textwidth,height=.30\textwidth]{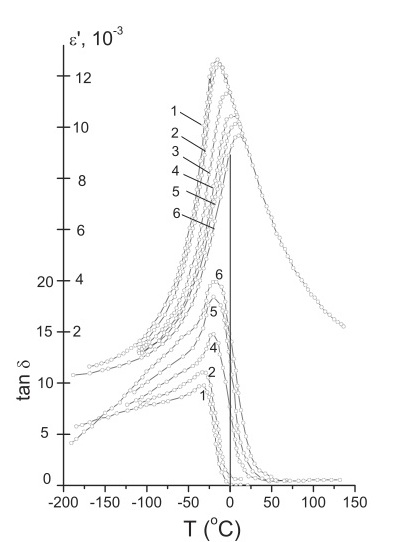}\hfill
\includegraphics*[width=.2950\textwidth,height=.30\textwidth]{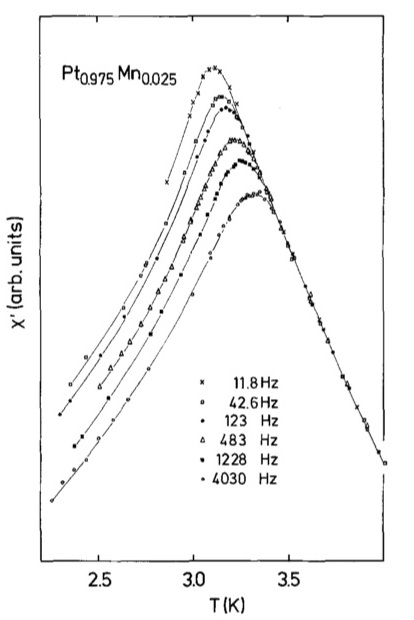}\hfill
\includegraphics*[width=.2950\textwidth,height=.30\textwidth]{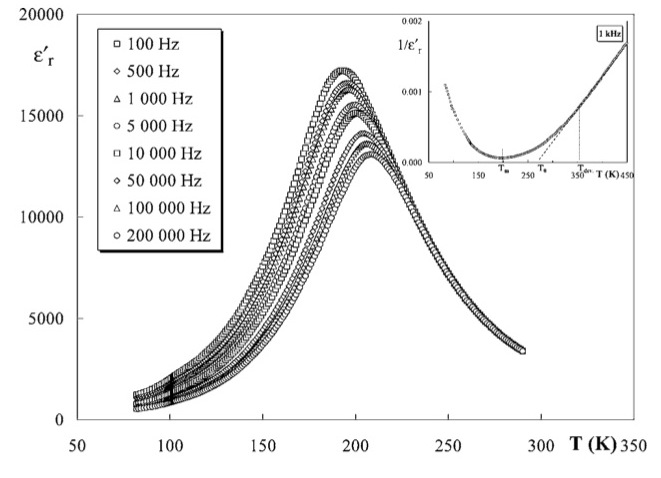}\hfill
\caption{Frequency-dependent AC susceptibilities;(i)  heterovalent relaxor  $\mathrm {Pb(Mg}_{1/3}\mathrm{Nb}_{2/3}\mathrm{)O}_3$ (PMN) \cite{Smolenskii}, \mbox{(ii) spin glass {\bf{Pt}}Mn \cite{Wassermann}  \copyright {Springer (1983)}}, (iii)
homovalent relaxor  $\mathrm{BaZr}_{0.35}\mathrm{Ti}_{0.65}\mathrm{O}_3$ (BZT) \cite{Maglione} \copyright{IOPP (2004)}; frequencies reducing right to left.
}
\label{fig:ac_susceptibilities}
\end{figure*}

 The first class of systems that I wish to note are the so-called  `relaxors'
 \footnote{Also known as 'relaxor ferroelectrics' \cite{Cross}.},
site-disordered displacive ferroelectrics, which experimentally exhibit  peaks in their measured dielectric susceptibilities with significant frequency-dependence, very reminiscent of the behaviour of the magnetic susceptibility of conventional spin glasses near their transition temperatures; illustrated in 
 Fig \ref{fig:ac_susceptibilities}. Also like in spin glasses, in the absence of driving fields there is no global polarization. In consequence of these similarities, it is natural to look for similar conceptual origins \cite{Viehland}. I shall argue that there are indeed such similarities but also that their consideration poses further issues and questions for statistical physics. 
 In fact, however, studies of relaxors and of spin glasses have  evolved  independently and have  largely remained so. The  observations in Fig \ref{fig:ac_susceptibilities}(i) date from the mid-1950s, almost two decades before the observation of the sharpening of a cusp in the low-field susceptibility of {\bf{Au}}Fe by Cannella and Mydosh \cite{Cannella-Mydosh} lit the theoretical explosion of interest in spin glasses. 
 
Spin glasses have received much study, both experimental and theoretical and are considered largely understood, albeit with some remaining controversies concerning issues such as the existence or otherwise of `replica-symmetry-breaking' and of  true spin-glass phase transitions in external fields and in systems of  finite dimensions;   the case of spin glasses with range-free interactions \cite{SK}  is considered solved \cite{MPV, Panchenko}.  It is generally recognised that the key ingredients for spin glass behaviour are frustrated interactions and effectively quenched disorder. However, the mechanism and essential ingredients for relaxor behaviour remain incompletely understood and disputed ({\it{e.g.}}\cite{Kleemann}). 
 
The topic of relaxors has been pursued mainly by experimentalists, including considerations of applicable materials design, and a smaller number of theorists mainly concerned with understanding and demonstrating  experimental observations via  (classical) computer simulations of the finite-temperature behaviour, involving many parameters calculated using realistically full (quantum) first-principles evaluation of model parameters. In contrast, in spin glasses the theoretical interest has been more in studying minimal models to understand  the basic physics of the spin glass state, the character and implications of its unconventional features,  the fundamental conditions for the existence of true spin glass transitions, and   generalizations and applications of the discovered subtle concepts and novel mathematical methods, mostly in topics far beyond the materials that  first  stimulated their study. Here my intent is to consider displacive relaxors from a similar minimal conceptual viewpoint and to relate to  issues of possible potential  interest to spin glass theorists, through observations and analogies rather than detailed calculations.

\begin{figure}[h]
\includegraphics[width=.5\textwidth,height=.30\textwidth]{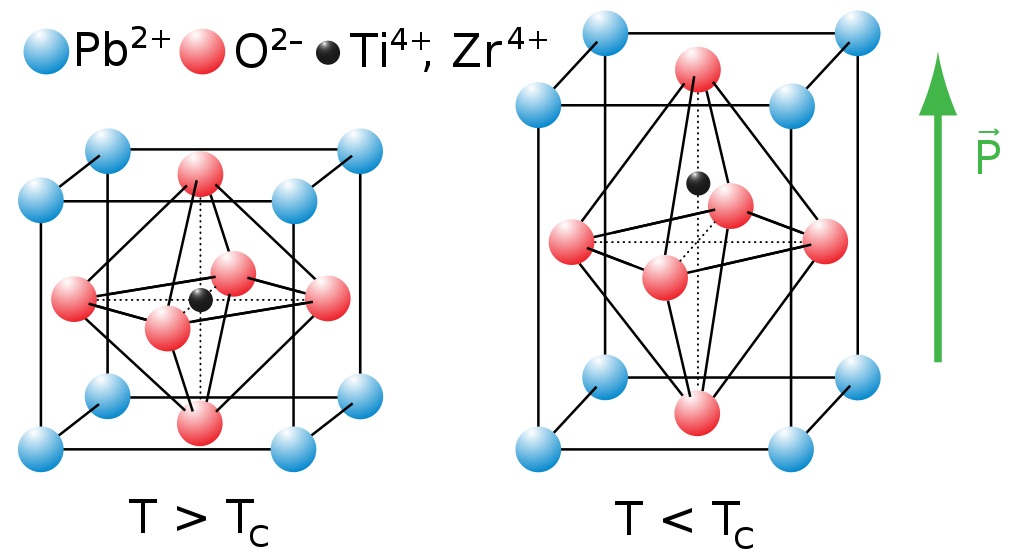} 
\includegraphics[width=.24\textwidth,height=.28\textwidth]{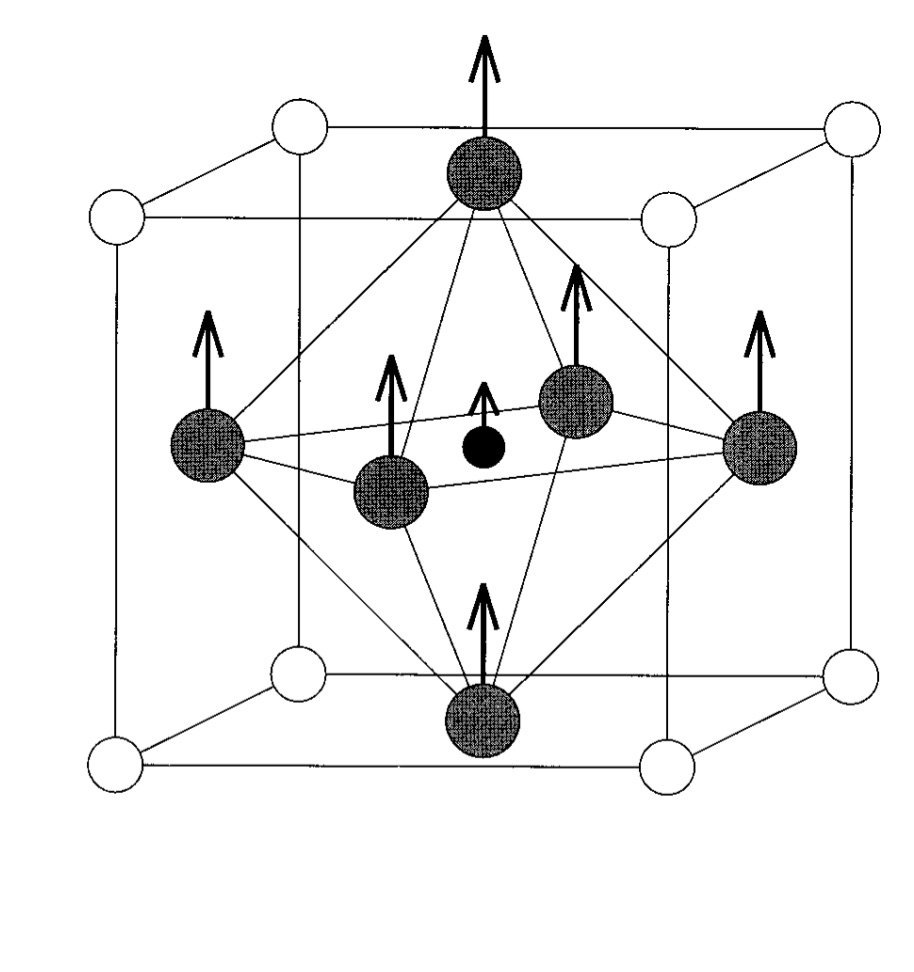}
\includegraphics[width=.24\textwidth,height=.26\textwidth]{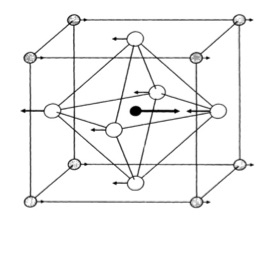}
\caption{Left half: Unit cell structure of $\mathrm {PbTiO}_3$ above and below the ferroelectric transition temperature.
 Right half: Relative ionic displacements in  tetragonal  (left) ${\mathrm{PbTiO}}_3$ \cite{Garcia-Vanderbilt} and (right)  ${\mathrm{BaTiO}}_3$ \cite{ZVR}.}
\label{fig:PbTiO3}
\end{figure}

Displacive ferroelectrics are ionic crystals which undergo ion-displacive phase transitions from higher-temperature high-symmetry structures without any global electric moments to lower-temperature phases of lower-symmetry with different intracell displacements of ions of opposite signs,  leading to overall electric dipolar moments. This is illustrated in Fig \ref{fig:PbTiO3}(left) for  $\mathrm {PbTiO}_3$,  which at higher temperature has the classic (cubic) perovskite structure and formula  ${\rm{ABO_3}}$ with nominal charges
 $\rm{A^{++}}$, $\rm{B^{++++}}$  
and $\rm{O^{ - -}}$, distorting as temperature is lowered beneath a critical temperature
 to a tetragonal structure with the ions also internally displaced relatively to one another so as to yield a ferroelectric moment, as further   illustrated  in Fig \ref{fig:PbTiO3}(right)  for $\mathrm {PbTiO}_3$  \cite{Garcia-Vanderbilt}  and   $\mathrm {BaTiO}_3$ \cite{ZVR}.
All three examples of Fig \ref{fig:ac_susceptibilities} involve the combination of (i) pairwise interactions that involve mutual frustration/competition but are essentially lattice-periodic and \mbox{(ii) quenched} atoms/ local disorder that works against a simple compromise of periodic order. In the spin glass of Fig \ref{fig:ac_susceptibilities}(ii) the relevant microscopic variables  are the orientations of local moments/fixed-length spins 
(on the {\bf{Mn}}), 
subject to long-range oscillatory/frustrated  Ruderman-Kittel-Kasuya-Yosida  (RKKY) interactions, $J\cos(2k_{F} R)/R^{3}$. In the two relaxor examples the relevant microscopic  entities are continuously-valued and oriented  displacements of ions from their  positions in an ideal perovskite structure, essentially soft {\it{pseudo}}-spins,  interacting via a combination of shortish-range quantum mechanical effects and long-range Coulomb/dipolar interactions. In all the examples the quenched disorder arises from the different types of atoms/ions occupying the nominal lattice sites.

The original and most famous displacive relaxor of Fig \ref{fig:ac_susceptibilities}(i) is an alloy $\mathrm {PbMg_{1/3}Nb_{2/3}O}_3$, commonly known as PMN, in which the $\rm{Ti^{++++}}$ ions of $\mathrm {PbTiO}_3$ (coloured black in Fig {\ref{fig:PbTiO3}})  
are replaced randomly and heterovalently by $\rm{Mg^{++}}$  and $\rm{Nb^{+++++}}$ in the ratio 1:2, thereby introducing disorder in  the site occupation, the local restoring forces and
 the effective interactions driving  correlations between the displacements at different sites,  along with effective random fields. 
 
\subsection{Homovalent alloys}

In considering the modelling  of relaxors it is, however,  easier (from a spin glass theory perspective) to start with  the homovalent alloy $\mathrm{BaZr}_{x}\mathrm{Ti}_{1-x}\mathrm{O}_3$ (BZT), an alloy of $\mathrm {BaTiO}_3$ and $\mathrm {BaZrO}_3$  in which $\rm{Ti^{++++}}$  and $\rm{Zr^{++++}}$ ions are randomly placed on the B-sites of the ${\rm{ABO}}_3$ structure; Fig \ref{fig:ac_susceptibilities}(iii) refers to an example with $x=0.35$.  
Because the Ti and Zr ions are of the same charge (4+), to a first approximation  there are no consequential extra charges or fields and the intersite  Coulomb interactions are relatively unaltered.
However, there is still important site disorder, with the Ti and Zr ions having different local 
displacement-energy coefficients. That for the term quadratic in the displacement of Ti ions is positive but small  enough that energy reduction due to interaction bootstrapping is sufficient so that $\mathrm {BaTiO}_3$ is a ferroelectric at low enough temperature, with coherent spontaneous ionic displacements from  pure perovskite. In contrast, the local harmonic restoring coefficient of  
Zr ions is too great for $\mathrm {BaZrO}_3$ to be able to bootstrap ferroelectricity via interaction gains. 
Hence, in BZT the Zr ions are analogues of the non-magnetic constituents in normal spin glass alloys, with relaxor  (or ferroelectric) behaviour arising from the Ti.

The Hamiltonian for a homovalent relaxor alloy 
can be expressed as \footnote{Strictly, there should also be a term corresponding to coupling to the global strain \cite{Zhong}, important to generate the (observed) structural feature of change of global unit cell 
accompanying a ferroelectric transition.
However,  for the relaxors it is observed that the global structure remains cubic. Hence, for the present (qualitative) discussion of relaxors  we shall instead assume that the strain coupling  can be absorbed into the effective parameters of eqn.(\ref{eq:Displacive}).}
\begin{equation}
\fl
H=\sum_{i} 
\{\kappa_{i}| \textit{\textbf{u}}_{i} |^2
+\lambda_{i}|\textit{\textbf{u}}|^4    
+ \gamma_{i}  
{{(u_{ix}^{2}u_{iy}^{2}
 + u_{iy}^{2}u_{iz}^{2} +u_{iz}^{2}u_{ix}^{2})  }}
\}
- {\sum_{(ij)}\sum_{\alpha \beta}J^{\alpha  \beta} (R_{ij})     u_{i\alpha}u_{j\beta}},  
\label{eq:Displacive}
\end{equation}
where the coefficients $\kappa$ \footnote{In much of the relaxor theory literature $\kappa$ is denoted $\kappa2$.}, $\lambda$ and $\gamma$ depend on the types of ions at the corresponding sites and the $\{\bf{u\}}$ are their local displacements from their positions in the perfect) perovskite \cite{Kingsmith, Zhong, Akbarzadeh}\footnote{First principles calculations on a range of pure systems indicate  $\lambda>0$ and $\gamma<0$.}.
The interaction term has short-range exchange and long-range frustrated Coulomb/dipole contributions, the latter going as
\begin{equation}
 H_{dip}= \sum_{(ij)}(Z_{i}Z_{j}/\epsilon)[{\bf{u}}_{i}.{\bf{u}}_{j}
 - 3 ( {\bf{\hat{R}}}_{ij}.{\bf{u}}_{i})( {\bf{\hat{R}}}_{ij}.{\bf{u}}_{j})]
 / |{\bf{R}}_{ij}|^3.
 \label{H_dip}
 \end{equation}

There is an immediately apparent analogy with a continuous  spin-length (soft)  version of the Hamiltonian for a conventional classical Heisenberg spin-glass
\begin{equation}
H=\sum_{occupied \, sites}
\{
\kappa{|\boldsymbol{\phi}}|_{i}^2
+\lambda{|\boldsymbol{\phi}_{i}|^4    
\}   
- \sum_{ij} J(R_{ij})  
{\boldsymbol{\phi}_{i} }.
{\boldsymbol{\phi}_{j}},  
}
\label{eq:Spin glass_continuous_spin}
\end{equation}
where the usual hard-spin S limit results from 
\begin{equation}
\kappa  \rightarrow -\infty, \lambda \rightarrow \infty, \kappa/2\lambda \rightarrow -S^2.
\label{Ising_limit}
\end{equation}
\setcounter{footnote}{0}
However, by contrast, in the displacive perovskites of interest $\kappa$ is  positive, so an isolated ion $i=0$ in an otherwise pinned structure $\{\phi_{j\neq 0}=0\}$ would not displace (no local moment). For  macroscopic ferroic order at sites with positive $\kappa$ the bootstrapped energy gain from interactions needs to be sufficient to overcome the local displacement cost, resulting in {\it{`induced moments' }} on the relevant sites \footnote{{\it{c.f}} itinerant magnets. See also subsection {\ref{Itinerant cluster glasses}}}.
In the case of $\mathrm {BaTiO}_3$, the $\kappa_{Ti}$ is smaller than the critical value for spontaneous coherent ferroelectric order, while for $\mathrm {BaZrO}_3$ $\kappa_{Zr}$ is too great and there is no ferroic order. In consequence one can deduce that any order on B-sites in BZT is driven by the Ti ions. Also, in $\mathrm {BaTiO}_3$ it is observed that of the positive ion displacements those of the Ti are  dominant \footnote{compared with the ${\rm{Ba^{++}}}$.}. Hence, it is natural to concentrate, in minimalist modelling, on the Ti ions
\footnote{Or on normal modes  centred on the Ti ions \cite{Akbarzadeh}.}.

Monte Carlo simulations \cite{Akbarzadeh} of a model of $\mathrm{BaZr}_{x}\mathrm{Ti}_{1-x}\mathrm{O}_3$ at $x=0.5$,  allowing for explicitly different local coefficients $\kappa$ for Ti and Zr but only averaged interactions $J(R)$,  have  demonstrated a `critical' temperature marking a peak in the quasi-ZFC susceptibility, evaluated from correlations using the fluctuation-dissipation relationship, together with the onset of a separation from the directly measured FC susceptibility, reminiscent of the corresponding behaviour in conventional spin glasses. Later molecular dynamics simulations \cite{Wang_NM2006} have yielded frequency-dependent susceptibility peaks similar to those measured experimentally and displayed in Fig {\ref{fig:ac_susceptibilities}(iii).

To the best knowledge of the present author, sophisticated studies of critical transitions via finite-size scaling of spin-glass correlation lengths, now-standard in spin glass simulations, have not yet been published for  homovalent relaxors \footnote{Yet-unpublished studies on simplified related  models have, however, indicated true pseudo-spin glass transitions (Andresen, private communication).},
although the conceptually analogous (discretized) Ghatak-Sherrington (GS) model \cite{GS}
\begin{equation}
H=\sum_{i}DS_{i}^2
 -\sum_{(ij)} J_{ij}S_{i}S_{j}; \ S=0,\pm1; \ {\rm{quenched \ random}}  \  \{J_{ij}\}
\label{GS}
\end{equation}
 has been studied by such methods  \cite{Leuzzi-Crisanti},   demonstrating the  induced moment spin-glass phase for small positive D (the analogue of small positive $\kappa$) .

\subsection{Heterovalent alloys}

Turning to the heterogenous case of PMN ($\mathrm {PbMg_{1/3}Nb_{2/3}O}_3$), 
one way of modelling, through extension from the spin glass perpective of  homovalent relaxors discussed above,  is to consider it as a fictitious alloy
$\mathrm{ PbMg^{*}_{1/3}Nb^{*}_{2/3}O_3} $, 
where Mg* and Nb* are fictitious ions with charges +4 but with all other model-relevant properties  the same  as those for $\rm{Mg}^{++}$ and 
$\rm{ Nb}^{+++++}$, along with extra charges -2 at Mg positions and +1 at Nb positions \cite{Sherrington_PMN}. Noting that the ionic radius of $\rm{ Mg^{++}}$ (86 pm.) is the same as that of $\rm{Zr^{++++ }}$ (86 pm.) and the ionic radius of $\rm{ Nb^{+++++}}$ (78 pm.) is  similar to that of $\rm{Ti^{++++}}$ (74.5 pm.)\footnote{The quoted (crystal) ionic radii are taken from a table on Wikipedia, itself collected from \cite{Shannon}.}, one might expect that $\mathrm{ PbMg^{*}_{1/3}Nb^{*}_{2/3}O_3} $ should behave like PZT ($\mathrm{PbZr}_{x}\mathrm{Ti}_{1-x}\mathrm{O}_3$) with $x=1/3$.

Given that BZT is a relaxor one might further naively anticipate that PZT might also be a relaxor, but this has not been observed
\footnote{PZT is, however,  of great application-value because of its significant piezoelectric properties, especially near a morphotropic phase boundary near $x=0.5$},
probably because Pb is much softer than Ba, as is observed in differences between the magnitudes of their displacements in the ferromagnetic phases of  $\rm{PbTiO_3}$ and $\rm{BaTiO_3}$  \footnote{ The soft optical modes responsible for the para-ferroelectric transition in  pure  $\rm{BaTiO_3}$ involve  approximately 4:1 Ti:Ba  but in $\rm{PbTiO_3}$ the corresponding split is 3:7 Ti:Pb \cite{Kingsmith}; see also Fig \ref{fig:PbTiO3}(right). } and as suggested by the differences in the ionic radii of $\rm{Pb^{++}}$ (133 pm.) and $\rm{Ba^{++}}$ (149 pm.).

Hence, one may  deduce that  $\mathrm{ PbMg^{*}_{1/3}Nb^{*}_{2/3}O_3} $ alone would probably not have a relaxor phase. So that leaves the likelihood that the random fields due the extra charges -2 (on Mg sites) and +1 (on the Nb sites) are also  needed to explain the relaxor behaviour of PMN, as was proposed in \cite{Westphal}.
Given that there is a common belief in the spin glass community that fields destroy spin glass transitions, this qualitative comparison already indicates  an intriguing situation for statistical physics.

PMN has also been simulated \cite{Bellaiche_PMN}, using a model with the dominant displacements on the Pb ions \footnote{This is in contrast with \cite{Akbarzadeh}, where the modes used were centred on the Ba ions.} 
and  a simply-averaged virtual-crystal interaction function, both (i) without and (ii) with  
 random fields due to the different extra effective charges on the ${\rm{Mg}}^{++} $ and 
 ${\rm{Nb^{+++++}}}$ B-ions.
In case (i) this demonstrated the sharp susceptibility peak expected of a ferroelectric, while case (ii) exhibited instead a more rounded peak at a lower temperature, interpreted as a relaxor peak and suggesting
   that random fields alone, without interaction disorder, might be capable of driving relaxor behaviour  in a system with frustrated interactions, although a more sophisticated simulational study would be required to be convincing of a true transition. 
   
\setcounter{footnote}{0}

 \subsection{Hard or soft (pseudo)spins.}
 
 As noted above, these relaxor ferroelectrics have positive (and usually small) values of $\kappa$.  Hence they can be considered as having `soft' (pseudo) spins, that require cooperative behaviour to become significant, in contrast to most systems conventionally studied in statistical physics, where the spins can be considered as `hard', much as in the difference between BCS (soft) and Bose-Einstein  condensation (hard) \footnote{in these cases of type II in the notation of \cite{Kohn-Sherrington}}. 
 There exist other hard dipolar and higher-moment analogues of hard-spin spin glasses \cite{Hochli}, but they are not pursued here.

\section{Strain Glass}

Another set of material systems  relevant to spin glass/ random field conceptualization are quenched disordered variants of the so-called `shape-memory alloy' Nitonol (NiTi), such as ${\rm{Ni}}_{0.5+x}{\rm{Ti}}_{0.5-x}$ and ${\rm{Ti}}_{0.5}{\rm{Ni}}_{0.5-x}{\rm{Fe}}_{x}$, 
which have recently been  observed to exhibit distortive pseudo-spin glass behaviour for sufficient $x$\cite{Sarkar, Wang}. NiTi itself is a metallic compound \footnote{In the literature, NiTi is usually referred to as an alloy, but  it is actually a chemical  compound with the two types of atoms arranged periodically.} which at higher temperature  is of simple cubic rocksalt (NaCl) structure, known as austenite, but which, as temperature is lowered,  undergoes a first-order distortive transition to a twinned phase of stripes of local cells of differently oriented tetragonal symmetry, known as martensite. Martensite can be thought of as the elastic analogue of  periodic ferro- or antiferro-magnetic order in magnetic systems and arising as a best compromise in a system with periodic but frustrated interaction but no quenched disorder. Disorderly alloying, for example randomly replacing some Ti with extra Ni or of some Ni by Fe, leads to the above-mentioned pseudo-spin-glass, now known as {\it{`strain glass' }} \cite{Sarkar, Ji}.

A relationship with  spin glass conceptualization is easily visualized by using  simple Ginzburg-Landau 
phenomenological modelling in terms of deviatoric strains.  For simplicity we consider a two-dimensional model where the local transition is from square to rectangle, characterized by the deviatoric strain $\phi = (\epsilon_{11} -\epsilon_{22})$, the $\epsilon_{\alpha \beta}$ being normal strain tensors. The local contribution to the free energy can then be modelled as
\begin{equation}
F_L= \sum_{i} [A_{i}(T){ \phi_{i}}^2 +B_{i}(T){\phi_{i}}^4 + C_{i}(T){\phi_{i}}^4],
\label{FL}
\end{equation}
where the \{i\} label local cells.
There are also effective site-to-site interactions
\begin{equation}
F_I=-\sum_{(ij)}   
J({\bf{R}}_{ij})  
 {\phi_{i}}    {\phi_{j}}
\label{FI}
\end{equation}
where $J(\bf{R})$ includes both the
usual short-range  ferromagnetic ($\nabla^2$) contribution  and a long-range term arising  from the  application of St. Venants constraints on the strains \cite{Lookman}, of the form 
\begin{equation}
J_{SV}({\bf{R}}_{ij}) \sim f(\cos{4\theta_{ij}})
  /  |{\bf{R}}_{ij}    |^2
\label{St_Venat}
\end{equation}
where $\theta_{ij}$ is the angle subtended by ${\bf{R}}_{ij}$ in the Cartesian coordinate frame of the lattice. This interaction is frustrated, with both positive and negative contributions as a function of $\cos{\theta}$ \footnote{Note that in \cite{Lookman} and subsequent papers by its authors,  the $\bf{k}$-space formulation of the interaction is used in numerical work.}. 

In a pure system, such as NiTi itself, there is no disorder among the local coefficients and the transition between austenite and martensite occurs when the temperature is lowered until A(T) is reduced (from a positive value) sufficiently until the lowest free energy occurs at a finite  $|\phi |$. 
However, in the case of local disorder in the $A_{i}$, combined with (even periodic)  frustration in the interactions, there arises the possibility of a spin-glass-like distortion in preference to the normal martensitic stripes \cite{Krumhansl}.

Noting that the austenite-martensite transition is observed as first order, it follows that B in eqn.(\ref{FL}) must be negative, C positive. It is thus natural to simplify further to a discrete
 formulation
\begin{equation}
F= \sum_{i} D_{i}(T)S_{i}^2 -\sum_{(ij)}J({\bf{R}}_{ij})S_{i}S_{j};  \  S=0,\pm1, 
\label{FS}
\end{equation}
where $S=0$ corresponds to a locally cubic cell and $S=\pm1$ correspond to the two possible orthogonal rectangular cells. With quenched disorder in the $\{D_{i}\}$ this yields a random site variant of the GS model and so can be expected to exhibit a spin glass phase for sufficient disorder as the mean D is reduced \cite{Sherrington_strain_glass}.

In the real alloys, one can anticipate that changes in chemical enviroments and point defects would introduce also  random  local terms linear in $\phi$ ( or $S$), alias random fields. Indeed, in the context of the earlier discussion of relaxors, it is interesting to note that phase-field computer simulations  of a model martensitic alloy including disorder purely of random field  character, but not random A , have been performed and argued to show  the same  sequence of phase transitions, austenite-martensite to austenite-strain glass as the quenched disorder is increased \cite{Ren_point_defects},  as is expected from the case of purely random A. Thus, this appears to suggest again that random field disorder, in combination with long-range  frustrated interactions but without site dilution or interaction disorder, may also be able to lead to a quasi-spin glass.

 \section{Coulomb Glass}
 
Another situation where spin glass-like behaviour driven by random field disorder has been proposed  is in the  Coulomb Glass problem \cite{Efros, Mueller}, which for half-filling can be expressed in pseudo-spin notation as an Ising model with long-range pairwise antiferromagnetic  Coulomb interactions, dependent only on the separation,  along with quenched random local fields. 
\begin{equation}
H=A\sum_{(ij)} \sigma_{i} \sigma_{j} / |{\bf{r}}_{ij}| +\sum_{i} h_{i}\sigma_{i}; \ \sigma=\pm 1; \  P(h) = P(-h)
\label{Coulomb}
\end{equation}
Analytic studies have demonstrated mean-field similarity to the SK spin glass and predicted a sharp phase transition to a spin-glass like phase above a critical disorder \cite{Mueller1, Mueller}.
A recent computer simulation of this case  \cite{Andresen_Coulomb_Glass} using modern sophisticated scaling tests has demonstrated a sharp transition from plasma/paramagnet to a Coulomb Glass state  as temperature is reduced, beyond a sufficient strength of quenched randomness \footnote{The indicative crossing of the simulated spin-glass correlation length normalized by the sample 'length' required a 4-replica evaluation procedure and was not observed in the usual 2-replica procedure.}.

Given that the antiferromagnetic Coulomb interaction  is frustrated, as are the interactions in the perovskite ferroelectrics above (eqn.(\ref{H_dip})) and the martensitic alloys (eqn.({\ref{FI}})), this observation gives extra weight to a speculation that in a system with appropriate periodic but frustrated interactions the addition of quenched disorder  through sufficiently potent  local random fields alone could induce a spin-glass/relaxor phase. What is `appropriate' is less clear; however, all these examples have interactions that have no cut-off and decay only as powers of $1/R$.

Beneath the critical disorder strength the low temperature phase of the Coulomb Glass problem is a periodic one, known as `charge-ordered' in reference to its character in the original electron occupation basis. It has its analogue in the perovskite displacement problem in  the ferroelectric phase found at levels of quenched disorder beneath a critical value, albeit that in the $\kappa > 0$ soft pseudo-spin cases sufficient potential binding energy is also needed to bootstrap moments at all.

\section{Polar nanoregions, tweed and cluster glasses}

\subsection{Relaxors}

A concept regularly employed in discussions  on relaxor ferroelectrics is of so-called polar nanoregions (PNRs),  clusters of  short-range correlated dipolar moments, onsetting well above the temperature of the relaxor susceptibility peaks and growing as temperature is lowered \cite{Cross, Egami}. Conceptually these can be expected as arising from statistical clusterings of ion placements in the nominal lattice, leading to a distribution of local displacement correlation strengths, with stronger regions visible to probes of appropriate timescales, as long as their temperature-dependant lifetimes are sufficient. 

A proper modelling of PNRs should involve a dynamical treatment, but as 
 a simple  guide/illustration 
one might consider a crude (but inhomogeneous) mean-field free-energy analogue
of the homovalent BZT model introduced above, eqn.({\ref{eq:Displacive}}), for simplicity considering only one-dimentional displacements:
\begin{equation}
F(T)=\sum_{i}  \{\kappa(T) <u_{i}>^2  +  \lambda_{i}(T)<u_{i}>^4  \}   
-  {\sum_{(ij)} J (R_{ij},T)    < u_{i}><u_{j}>  }, 
\label{eq:One-dim Displacive}
\end{equation}
where the \{$<u_{i}>$\} represent effective local-moment averages  of the \{{$u_{i}$\}}.
Minimising with respect to the $\{<u_{i}>\}$ yields the self-consistency equation
\begin{equation}
\kappa_{i}(T)<u_{i }>    
- \sum_{ j\neq {i}}J(R_{ij},T)<u_{j}> 
=- 2\lambda_{i}(T)<u_{i}>^3.
\label{mfsc}
\end{equation}
Clearly, there are always solutions $\{<u>=0\}$. However,
for the \{$\lambda$\} positive \footnote{as computed for T=0 \cite{Kingsmith}}, comparison with the eigenequation
\begin{equation}
\kappa_{i}(T)v_{i }    - \sum_{ j\neq {i}}J(R_{ij},T)v_{j} =Ev_{i}
\label{eigenequation}
\end{equation}
suggests that solutions of eqn.(\ref{mfsc}) with $<u>\neq 0$ 
require that 
$A_{ij} \equiv\{ {\kappa_{i}(T)\delta_{ij}} -J(R_{ij},T)\}$ 
has negative eigenvalues.  The most relevant $T$-dependence is anticipated for the $\{\kappa(T)\}$ which are expected to increase as $T$ increases, so that at high temperature the system is paraelectric, but with temperature reduction allowing for a lower value yielding an ordered phase if sufficient.

\setcounter{footnote}{0}

For the pure case (all $\kappa$ equal) all these eigenfunctions are extended, with the ferroelectric phase transition signalled by the band edge 
reaching $E=0$ as $T$ is reduced.  However,  for a system with quenched disorder it can be anticipated that the solutions of eqn. (\ref{eigenequation}) near the band edge will  have finite lifetimes,  unstable against fluctuations and effectively localized, with the true phase transition  delayed until a lower temperature, depending on the degree of quenched disorder
\footnote{{\it{c.f.}} localization edge}.

Note that this argument for  finite-lifetime short-range correlated clusters above a true phase transition extends to (disordered) alloys with either  spin-glass-like or quasi-periodic  ordered phases, with the upper temperature  limit of PNRs of the order of the ferroelectric transition of the more ferroelectric pure state \footnote{Strictly, without the global strain modification.}}
\footnote{{\it{c.f.}} Griffiths phases \cite{Griffiths}}.
Thus for homovalent alloys one might expect a phase diagram of the schematic form of Fig.\ref{Schematic_with_PNRs}. 
\footnote{An alternative crude approximate/qualitative way to argue for such a schematic phase diagram 
would  be to consider the minima of eqn(\ref{eq:Displacive}) in which all sites with a local energy less than order $T$ are frozen out (put to $\phi =0$). At high $T$ all sites would have $\phi =0$. The appearance of finite clusters with $\phi \neq 0$ at $T$ is reduced would indicate the onset of PNRs, while the onset of percolation of sites with $\phi \neq 0$ would indicate the transition to order, ferroelectric or relaxor.
}
\begin{SCfigure}
\includegraphics[width=.50\textwidth,height=.45\textwidth]{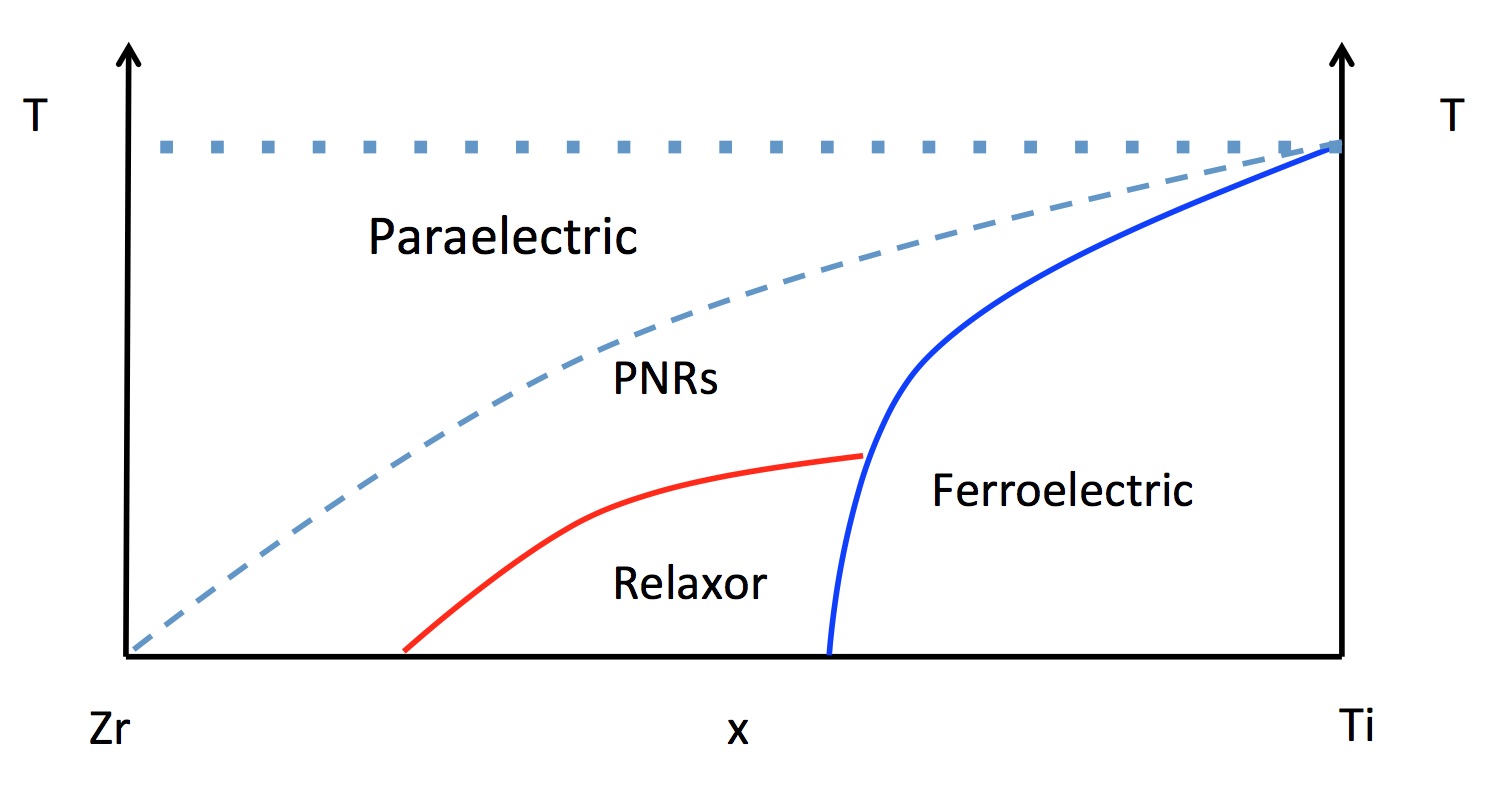}
\caption{Schematic `phase diagram' for  BZT expected in the light of heuristic considerations; from \cite{Sherrington_Lookman}. Solid lines denote true phase transitions. The dotted line indicates the onset of PNRs in the picture discussed.
The dashed line is a speculative illustration of crossover for the onset of significant visibility of PRNs. 
\vspace{1.1  cm}}
\label{Schematic_with_PNRs}
\end{SCfigure}

For heterovalent alloys account must also be taken of the random fields, so that  extra terms of the form 
$\sum_{i} {\bf{h}}_{i}.{\bf{u}}_{i}$ are needed in the Hamiltonian and consequently in an effective mean-field theory. Also, as noted earlier, it appears that the most important positive ions with regard to displacements in Pb systems are the A-site ions rather than the B-site ones on which the extra charges reside. Even without the interaction effects integral to $\rm{PM^*N^*}$ the extra random charges on the B-sites of PMN would lead to effective random fields on the Pb ions such as to lead to corresponding thermodynamically weighted random displacements of the Pb ions from their sites in the pure perovskite \cite{Sherrington_PMN}  extending even to high  temperatures.
Thus, different behaviour can be anticipated for the temperature extent of PNRs in homogeneous and heterogeneous alloys.

It should be noted, however, that there remains disagreement among practitioners on the character and origin of PNRs, in both homovalent and heterovalent relaxor systems.  

\subsection{Martensitic alloys}

In the case of martensitic materials, already for many years a pre-martensitic `phase' exhibiting intermixed domains of different martensitic orientations and  of austenite,  known as {\it{`tweed',}} was observed for a range of temperatures above the transition to the martensitic striped phase. It has now been interpreted as an analogue of the PNRs discussed above for ferroelectric alloys \cite{Zhang_2}, separate from strain glass \cite{Ren_tweed}.

\subsection{Itinerant cluster glasses}
\label{Itinerant cluster glasses}

Most statistical physics studies of magnetism consider local moment spins of discrete length. However, it has been known for a long time 
\cite{Stoner}
 that one can have itinerant ferromagnetism induced due to collective behaviour of conduction electrons  in transition metals \cite{Morosan}. 
Already in the 1970s,  this concept was extended to an itinerant analogue of local moment spin glasses \cite{Mihill, Mathon, 
Hertz}, but has not received much further consideration\footnote{But see e.g. \cite{Sachdev}.}.   

Theoretical modelling  of disordered transition metal alloy magnetism can be expressed so as to lead to close analogy with the discussion of relaxors presented above  \cite{Mihill, DSPSSB}. 
Let us start with a simple random Hubbard model for a transition-metal alloy of two metals, A and B, of different on-site Coulomb repulsion strengths 
\begin{equation}
H_{HA}=\sum_{ij;s=\uparrow,\downarrow} t_{ij} a_{is}^\dagger a_{js} +\sum_{i;s=\uparrow,\downarrow}V_{i} a_{is}^{\dagger}a_{is} + \sum_{i}U_{i}\hat{n}_{i\uparrow}\hat{n}_{i\downarrow},
\label{eq:Hubbard}
\end{equation}
where the $a,a^{\dagger}$ are site-labelled d-electron annihilation and creation operators, $\hat{n}_{is}=a^{\dagger}_{is}a_{is}$, and in general the $t_{ij}$, $V_{i}$ and $U_{i}$ depend upon 
the type of atoms at sites {\it{i,j}}. Re-writing in terms of complete squares of local magnetization and charge fluctuation operators,
concentrating on the spin fluctuations and, for conceptual simplicity,  absorbing  effects of charge fluctuations approximately into average hopping and local terms, we consider as a minimal electronic model
\begin{equation}
H=\sum_{ij,s=\uparrow,\downarrow}t_{ij}a_{is}^\dagger a_{js} -{\frac{1}{4}}\sum_{i}U_{i}{\textit{\textbf{S}}}_{i}.{\textit{\textbf{S}}}_{i}
\label{Hubbard_S}
\end{equation}
where
\begin{equation}
{\textit{\textbf{S}}}_{i}
=\sum_{s=\uparrow,\downarrow}   
 {a_{is}^{\dagger}}
{\boldsymbol{\sigma}}_{s,s'}a_{is'}.
\end{equation}

Using a Lagrangian functional integral many-body formulation, 
an `inverse completion of a square' procedure
\cite{Gelfand} 
\footnote{{\it{c.f.}}\cite{Stratonovich, Hubbard}}
may be employed  to effectively linearize the Hamiltonian in $ a_{is}^\dagger a_{js'}$  through the introduction of an auxiliary magnetization field variable $\textit{\textbf{m}}$, conjugate to ${\textit{\textbf{S}}}$ \cite{Sherrington_Aux}. One can then further `integrate out'  the original electron operators in favour of a description in terms purely of auxiliary magnetization variables.  Further taking the static approximation yields  
an effective free energy functional in local magnetization variables; to fourth order,
\noindent
\begin{minipage}{\linewidth}
\begin{eqnarray}
&F_{m}= \sum_{i}  (1-U_{i}\chi_{ii})   
|{\textit{\textbf{m}}}_{i}|^2 - 
 \sum_{ij;i\neq j}U_{i}^{1/2}U_{j}^{1/2}\chi_{ij}{\textit{\textbf{m}}}_{i}.{\textit{\textbf{m}}}_{j} \nonumber \\
 &- 
\displaystyle{\sum_{ijkl;\alpha\beta\gamma\delta}}(U_{i}U_{j}U_{k}U_{l})^{1/2}\Pi^{\alpha\beta\gamma\delta}_{ijkl}
m^{\alpha}_{i}m^{\beta}_{j}m^{\gamma}_{k}m^{\delta}_{l} + {\rm{higher \ order \  terms}}
\label{H_Sm}
\end{eqnarray}
\end{minipage}
where $\chi$ is the static band susceptibility function of the bare system, and $\Pi$ is a corresponding  bare 4-point function.
 A further change of variables
\begin{equation}
{\textit{\textbf{M}}}_{i}=U_{i}{\textit{\textbf{m}}}_{i}
\end{equation}
immediately brings this to the form  of Eqn.({\ref{eq:Displacive}}):
\begin{eqnarray}
&F_{M}= \sum_{i}  (U_{i}^{-1}-\chi_{ii})   
|{\textit{\textbf{M}}}_{i}|^2 - 
 \sum_{ij;i\neq j}
 \chi_{ij}{\textit{\textbf{M}}}_{i}.{\textit{\textbf{M}}}_{j} \nonumber \\
 &- 
\displaystyle{\sum_{ijkl;\alpha\beta\gamma\delta}}
\Pi^{\alpha\beta\gamma\delta}_{ijkl}
M^{\alpha}_{i}M^{\beta}_{j}M^{\gamma}_{k}M^{\delta}_{l},
\label{H_SM}
\end{eqnarray}
with local self-energy disorder weight $(U_{i}^{-1}-\chi_{ii})$ the analogue of  $\kappa$ in Eqn.({\ref{eq:Displacive}}) 
and 
$\chi_{ij}$
the analogue of $J_{ij}$.

Before considering finite-concentration alloys, we note that simple consideration of a system with two components A and B with $U_{A}=0$ but $U_{B }> 0$ immediately yields the well-known mean field results: (i) 
pure A is paramagnetic; (ii) pure B is ferromagnetic  if the Stoner criterion $({1-U_{B}\sum_{j}\chi_{ij})} \equiv ({1-U_{B}\chi(q=0)})<0$ is satisfied, but otherwise paramagnetic, (iii)
an isolated B  in an A-host will only carry a moment if 
$({1-U_{B}\chi_{ii})} \equiv ({1-U_{B}\int_{q}\chi(q))}<0$,  the Anderson condition \cite{Anderson_1961}  \footnote{The analogues for the relaxors are (ii)  Stoner criterion (ferroelectric): $\kappa -\sum_{j} J_{ij} <0$  and (iii) Anderson condition (local moment):  $\kappa <0$.}

 In metals $\chi_{ij}$ oscillates in sign as a function of separation, so  the effective interactions of Eqn. ({\ref{H_SM}}) are frustrated. It also has long-range decay as an inverse power of R.  Hence,   an ${\rm{A_{1-x}B_{x}}}$ alloy with a sufficient concentration $x$ of $B$ atoms would be expected to exhibit  the sequence of behaviours {{(Pauli) paramagnet - itinerant spin glass - ferromagnet}} as the concentration of B is increased from 0 to 1 \cite{Mihill},  analagously to the corresponding sequence predicted for a homovalent relaxor and depicted qualitatively  in  Fig \ref{Schematic_with_PNRs}. This phase structure was already reported in experiments on {\bf{Rh}}Co in the early 70s \cite{Coles-Tari}, but again further studies of this system seem not to have been pursued.
 
Concerning analogues of  PNRs, clearly the same arguments as used above for homovalent relaxors can be applied to the transition metal alloys.  In fact, already in 1973 this locally-coherent cluster concept was proposed for alloys like {\bf{Rh}}Co, formalized using the mapping to the effective magnetization field description of eqn.(\ref{H_SM}) and its further minimization and comparison with the localization eigenequation ({\ref{eigenequation}}),   along with the further suggestion that residual interactions between the clusters might lead to a  spin glass.\cite{Mihill}.  The lifetime of isolated clusters in the itinerant magnet case are, however, expected to be much shorter than for the displacive problem.

\section{Conclusion}

Several materials systems of active interest exhibiting ferroic glassy behaviour have been considered in analogy with spin glasses, with aims  both to understand the materials systems and to suggest issues for statistical physics beyond those of conventional spin glass studies. These involve both effective random interactions and random fields.

We have argued for many analogues of spin glasses involving soft induced-moment pseudo-spins, in contrast to the most-studied hard spins of conventional spin glass physics, including extending the concept of itinerant periodic magnetism to intinerant spin glasses.

Examination and comparison of measured properties in materials systems and observations of some computer simulations of theoretical models has led to an apparent conclusion that 
{\it{either}} random dilution {\it{or}} random fields have the potential to yield qualitatively similar glassy features in many systems with frustrated long-range interactions, both for systems with  good local moments already in their higher temperature para phases and for others in which moments need to be cooperatively induced in ordered phases. Spin glass-like  behaviour  in the case of random dilution of a system with frustrated interactions is expected and believed to be conceptually similar to that of random bond disorder \cite{Edwards-Anderson}, but the existence of true transitions in short-range disordered and frustrated systems in the presence of fields is still hotly contested, while the conventional (ferromagnetically interacting) random-field Ising model has been argued to not permit a spin glass thermodynamic phase \cite{Krzakala}. Thus, questions remain, on the correct vision of some of the materials systems discussed above ({\it{ e.g.}} whether the observed diffuse susceptibility peaks  indeed correspond to  true phase transitions in the thermodynamic static limit)  and on conditions on interaction forms for the possible existence of spin glass behaviour  (i) where the only quenched disorder is of local random fields and (ii) where there are both effective bond disorder and random fields. 

Attention has also been drawn to observations of  finite-lifetime pseudospin-correlated clusters in several disordered  ferroelectric and ferroelastic materials, as well as predicted for some transition metal alloys, and a crude consideration has been presented in terms of a na{\"i}ve inhomogeneous mean field modelling, but more sophisticated analysis and simulation, including dynamics, would be useful for better understanding. In principle one could imagine mapping to a basis 
of  eigenstates of eqn.(\ref{eq:Displacive}) with interactions between them, developing with temperature, and attempting to emulate and build on the qualitative conceptualizations that have been expressed, but currently a formal expression of this is missing.

 Noting that effective random fields occur naturally in the heterovalent displacive ferroelectric alloys, in contrast to normal magnetic alloys, and  they are not collinear, unlike the effective fields in gauge-transformed diluted antiferromagnets \cite{Aharony-Fishman}, it seems possible that some ferroelectric and ferroelastic alloy materials systems could provide experimental testbeds complementary to those available in magnetic systems.

 \section{Acknowledgements}
I would like to thank my now-deceased friend and colleague Roger Cowley for stimulating my interest in ferroelectric relaxors and Wolfgang Kleemann, Rasa Pirc, Laurent Bellaiche and Peter Gehring  for useful discussions and correspondence on the topic; Turab Lookman and Avadh Saxena for stimulating my interest in martensitic alloys and Xiaobing Ren for further valuable discussions; and Helmut Katzgraber, Juan Carlos Andresen and Moshe Schecter for interactions concerning computer  simulations.

Concerning spin glasses, I am very grateful to many friends and colleagues who have influenced my thinking, starting with Bryan Coles, Sam Edwards, Phil Anderson, Scott Kirkpatrick, Mike Moore, Alan Bray, David Thouless, Nicolas Sourlas and Giorgio Parisi, but also including many others, mostly part of the statistical physics community that grew out of the European networks that Giorgio, Nicolas and I initiated more than three decades ago, too many to list individually but all deeply appreciated.

\end{document}